**Graphene-Silicon Layered Structures on Single-crystalline Ir(111) Thin Films**

Yande Que, Yong Zhang, Yeliang Wang, Li Huang, Wenyan Xu, Jing Tao, Lijun Wu, Yimei Zhu, Kisslinger Kim, Michael Weinl, Matthias Schreck, Chengmin Shen, Shixuan Du, Yunqi Liu, and H.-J. Gao[*]

Y. D. Que,[†] Y. Zhang,[†] Prof. Y. L. Wang,[†] Dr. L. Huang, Dr. W. Y. Xu, Prof. C. M. Shen, Prof. S. X. Du, Prof. H.-J. Gao
Institute of Physics & University of Chinese Academy of Science, Chinese Academy of Science, Beijing, 100190, China
E-mail: hjgao@iphy.ac.cn
[†] These authors contributed equally to this work.

Dr. L. Huang, Prof. Y. Q. Liu
Institute of Chemistry, Chinese Academy of Science, Beijing, 100190, China

Prof. J. Tao, Dr. L. J. Wu, Prof. Y. M. Zhu
Condensed Matter Physics and Materials Science Department, Brookhaven National Laboratory, Upton, New York 11973, USA

Dr. K. Kim
Center for Functional Nanomaterials, Brookhaven National Laboratory, Upton, New York 11973, USA

M. Weinl, Prof. M. Schreck
Universität Augsburg, Institut für Physik, D-86135 Augsburg, Germany

## 1. Introduction

Epitaxial growth of graphene on transition metal crystals, such as Ru,[1-3] Ir,[4-6] and Ni,[7] provides large-area, uniform graphene layers with controllable defect density, which is crucial for practical applications in future devices. To decrease the high cost of single-crystalline metal bulks, single-crystalline metal films are strongly suggested as the substrates for epitaxial growth of large-scale high-quality graphene.[8-10] Moreover, in order to weaken the interactions of graphene with its metal host, which may result in a suppression of the intrinsic properties of graphene,[11,12] the method of element intercalation of semiconductors at the interface between an epitaxial graphene layer and a transition metal substrate has been successfully realized.[13-16] In particular, intercalation of a silicon layer has aroused a great interest in this research direction. It has been proven that uniform Si layers can be intercalated at the interface of large-area high-quality graphene and single-crystalline Ru and Ir bulk substrates, leading to recovery of the intrinsic electronic behavior of graphene, while the high quality of graphene is essentially intact after the Si intercalation.[13,14] Such an approach produces graphene-silicon hetero-layered structures compatible with the current Si-based microelectronic techniques and is an obvious big step forward toward the fabrication and integration of graphene-based devices. Taking advantage of these techniques, the low cost preparation of graphene-silicon hetero-layered structures on single-crystalline metal films rather than on metal bulks is highly promising for applications in future microelectronic devices. However, the formation and interfacial properties of such heterostructures has rarely been reported.

For the present work, we grew graphene on single-crystalline Ir(111) thin films, which were pre-fabricated on Si(111) substrate with an yttria-stabilized zirconia buffer layer – Ir(111)/YSZ/Si(111). Subsequently a silicon layer was successfully intercalated between the graphene and the Ir films. Low energy electron diffraction (LEED) measurements show that an ordered Si layer formed at the interface between the graphene and the Ir thin films. X-ray photoelectron spectroscopy (XPS) measurements reveal that the interfacial Si is in the form of pristine silicon rather than silicide. The details of such graphene-silicon hetero-layered structures are further revealed by high-resolution cross-sectional transmission electron microscopy (TEM), high-angle annular dark-field scanning transmission electron microscopy (HAADF-STEM) imaging and electron energy loss spectroscopy (EELS) analysis, together demonstrating that the intercalated Si-layer at the interface is about 5 Å thick. Raman and XPS spectroscopy measurements prove the decoupling of the graphene layer from the Ir(111) thin films and the recovery of intrinsic properties of graphene after Si intercalation.

## 2. Results and discussion

**Figure 1** schematically illustrates the formation of graphene-silicon heterostructures on single-crystalline Ir(111) thin films on Si(111) substrate. Highly ordered, large-area monolayer graphene was epitaxially grown on single-crystalline Ir(111) thin films (see experimental section for details). The formation and structure characterization of such epitaxial graphene was testified by LEED. Formation of graphene-silicon heterostructures was achieved via Si-intercalation approach. The quality of the Ir(111) films was checked by TEM, atomic force microscopy (AFM) and LEED. The Si intercalated samples were analyzed by LEED measurements. After Si intercalation, the electronic properties of graphene were checked by Raman and XPS spectroscopy. The interfacial Si layer was confirmed by XPS, high-resolution STEM and EELS.

### 2.1 Characteristics of single-crystalline Ir(111) thin films

**Figure 2**a is a cross-section TEM image of the starting sample of Ir(111)/YSZ/Si(111), which shows that the Ir thin films are very uniform, with a thickness of ~100 nm. High-resolution TEM image (Figure 2b) and corresponding fast Fourier transform (FFT) [inset of Figure 2b) reveal that the Ir films are well crystallized. Prior to surface cleaning of the Ir(111) thin films, AFM measurements unveil the presence of small pin-holes on the surface (Figure 2c). But such structural defects disappear after repeated cycles of argon ion sputtering and annealing in ultrahigh vacuum. The as-prepared Ir thin films exhibit a sharp LEED pattern with threefold symmetry (Figure 2d), indicating a highly ordered and (111) faceted surface.

### 2.2 Formation of graphene on Ir(111) films

Using such clean single-crystalline Ir(111) thin films as substrates, graphene was epitaxially grown via pyrolysis of ethylene. The as-prepared graphene on Ir(111) thin films exhibits a sharp threefold symmetric LEED pattern, as made clear in **Figure 3**a. Two sets of bright spots can be identified, as indicated by the white solid-line and dashed-line arrows. The outer set of spots corresponds to the reciprocal lattice of graphene, while the inner set of spots are assigned to the Ir(111) lattice. The parallel alignment of the two sets of spots suggests that the lattices of graphene and Ir(111) are parallel without any twisting.[6,17] The additional satellite spots result from the graphene moiré superstructure, which forms due to the lattice mismatch between graphene and Ir(111).[4] We note that individual LEED patterns obtained at different locations across the surface remain unchanged, confirming large dimension and high quality of the single-crystalline graphene layer on the substrate.

**2.3 Formation and characterization of graphene-silicon heterostructures**

To intercalate Si atoms at the interface of graphene/Ir, the as-prepared graphene was exposed to Si at RT followed by annealing at 800 K. Figure 3b shows an LEED pattern of the sample after Si intercalation. In addition to the spots originating from graphene and Ir(111), new spots appear, which are assigned to the two mirror domains of the √19 × √19 - R23.4 ° superstructure with respect to the Ir(111) surface. Meanwhile, the spots corresponding to the graphene moiré pattern become weaker, indicating a flattening of the graphene. Figure 3c shows the intensity profiles of LEED patterns across the solid lines indicated in Figures 3a (blue) and 3b (red) before and after Si intercalation. The intensity of spots for graphene remains nearly unchanged, whereas the spots for Ir decrease by a factor of ~3. Such changes of LEED patterns imply that the graphene sheet remains on the surface, and silicon atoms form an ordered layer at the interface. These behaviors are very similar to our previous work on the intercalation of a Si layer at the interface of graphene and Ir(111) single crystal.[14]

The properties of graphene in graphene-silicon heterostructures on Ir(111)/YSZ/Si(111) was also characterized with Raman and XPS spectroscopy, as shown in **Figures 4**a and b. Before Si intercalation, the Raman spectrum of graphene (red line in Figure 4a) is suppressed, which is due to a hybridization between the π bands of graphene and the Ir substrate, in line with previous studies.[14,18,19] After Si intercalation, in contrast, the two characteristic Raman features of graphene (black line in Figure 4a), the G peak (1616 cm$^{-1}$) and 2D peak (2712 cm$^{-1}$), are more clearly visible, suggesting that the intercalated Si layer effectively decouples graphene from the single-crystalline Ir(111) thin films and restores the intrinsic properties of graphene. This behavior is in agreement with our previous work on the intercalated Si layer at the interface of graphene and Ir(111) single crystal.[14] In addition, we observed blue shifts of both G and 2D peaks with respect to that of pristine graphene, indicating a slight residual strain in graphene and charge transfer from the underlying Si layer.[20,21] Figure 4b shows XPS spectra of C 1s centered around 283.8 eV for graphene before and after the formation of graphene-silicon heterstructures, the characteristic features of sp$^2$-hybridized C. The slight down shift (~0.2 eV) of the peak implies weaker charge transfer from the substrate to the graphene sheets.

Furthermore, the interfacial silicon layer between epitaxial graphene and Ir substrate was revealed via XPS spectroscopy measurements, cross-sectional TEM/STEM imaging and EELS analysis. XPS spectra of Si 2p are shown Figure 4c. Before Si intercalation, the sample is clear of silicon. After formation of graphene-silicon hetersostructures, it shows two dominate peaks centered at 99.7 eV and 102.0 eV respectively. The peak at 102.0 eV can be attributed to the formation of silicon oxide where silicon was not covered by graphene or on top of graphene. The oxidization of Si could occur when exposure to air during transfer the sample to the ultra-high vacuum system for XPS measurements. The peak at 99.7 eV can be assigned to pristine Si, suggesting that the interfacial Si beneath graphene is pristine silicon rather than Ir silicide. The presence of graphene avoid oxidization of the interfacial Si when exposure to air.

**Figures 5**a and **6**a show high-resolution cross-sectional TEM and STEM images, respectively, of the interface of the Si-intercalated graphene and Ir substrate. The highly-ordered structures of the Ir films suggest that the Ir film substrate remains stable and of high quality after the growth of the graphene and subsequent intercalation of the Si layer. Since the atomic number of carbon is much lower than that of Ir, the graphene layer that was already confirmed by the above mentioned LEED measurements cannot be distinguished in the STEM image, in which the contrast shows heavy atoms much brighter than light atoms.

Though the interfacial Si layer cannot be clearly seen in the STEM images, it was probed by EELS. Figure 5b shows the EELS spectrum around Ir-M edge, taken of the Ir(111) thin film, showing a broad peak around 2.2 keV. Whereas Figures 5c and 5d show the EELS spectra around Si L edge and K edge taken at the interface, respectively, identifying the existence of silicon atoms at the interface, the Si-$L_{2,3}$ EELS shows two main peaks at 104.5 eV and 125.9 eV, suggesting that Si may be oxidized to silicon oxide,[22] which could occur after exposure of the sample to air during TEM sample preparation. Moreover, an EELS line scan across the interface, with the intensity plot of Ir and Si quantified using the Ir-M edge and the Si-K edge as a function of position shown in Figure 6b, clearly demonstrates the distributions of Ir and Si atoms across the interface. The Ir intensity drops dramatically at the interface, indicating a sharp interface of the Ir thin films, while the Si intensity with much lower value changes within a thickness of ~ 0.5 nm, revealing Si existence in the form of monolayer or double layers at the interface.

## 3. Conclusions

We report here the formation of graphene-silicon hetero-layered structures on single-crystalline Ir(111) thin films and the corresponding geometric and electronic properties. Using Ir(111)/YSZ/Si(111) as the substrate, graphene with good quality was fabricated. Subsequent silicon intercalation led to the decoupling of the graphene from the Ir(111) thin films and the recovery of the intrinsic properties of graphene. The low-cost preparation of large-area high-quality graphene with intercalated Si layers at the interface of graphene and Ir(111)/YSZ/Si(111) substrate, is a promising route for graphene applications in nanodevices. Moreover, the graphene-silicon layered 2D structure on Si substrate takes advantage of current Si-based microelectronic techniques of device fabrication and integration.

## 4. Experimental Section

The growth of single-crystalline Ir(111) thin films has been well described elsewhere.[10,23] Sample preparation of graphene-silicon hetero-layered structures was carried out in an ultra-high vacuum (UHV) system with a base pressure of about $5 \times 10^{-10}$ mbar, equipped with standard surface processing and characterization facilities. First, the Ir(111) surface was cleaned by repeated cycles of sputtering with argon ions and annealing at 1200 K. High quality graphene was then fabricated via thermal decomposition of ethylene on as-prepared Ir(111) thin films. Silicon atoms were subsequently evaporated onto the graphene surface by electron-beam evaporation, and then the sample was annealed. LEED was employed to identify the superstructures macroscopically. XPS spectra were acquired via an ESCALAB 250 Xi XPS microscope using Al Kα X-ray source. Raman spectra were acquired by a Renishaw spectrometer at 532 nm with about 1 mW power. TEM samples were prepared using a Focused Ion Beam (FIB) system followed by ion-milling. TEM and STEM imaging and EELS analysis were performed on a double Cs-corrected JEOL JEM-ARM200F TEM with a Cold Field

Emission Gun and a Gatan Imaging Filter (GIF) Quantum system for EELS acquisition. Only the typical TEM/STEM/EELS results are shown here.

**Acknowledgements**

The work was financially supported by grants from MOST (Nos. 2013CBA01600, 2011CB932700, and 2011CB921702), NSFC (Nos. 61390501 and 61222112), and Chinese Academy of Sciences (Nos. 1731300500015 and XDB07030100) in China. The work done at Brookhaven National Laboratory was supported by the DOE BES, by the Materials Sciences and Engineering Division under contract DE-AC02-98CH10886, and through the use of the Center for Functional Nanomaterials.

Received: ((will be filled in by the editorial staff))
Revised: ((will be filled in by the editorial staff))
Published online: ((will be filled in by the editorial staff))

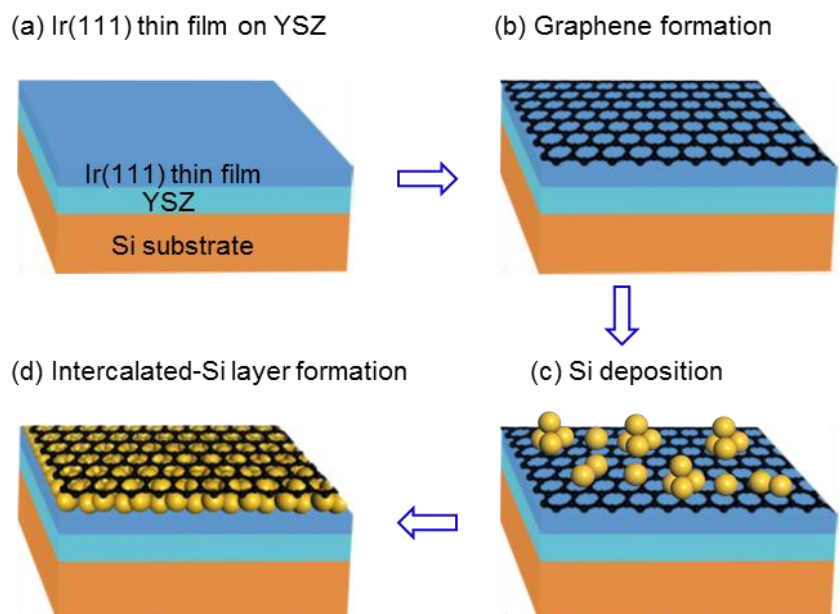

**Figure 1.** Schematic of the formation process of graphene-silicon layered structures. (a) Clean single-crystalline Ir(111) thin films on YSZ/Si(111) as supported substrates for graphene growth. Ethylene ($C_2H_4$) atoms are adsorbed on Ir(111) surface and decomposed at high temperature, leading to (b) the formation of epitaxial graphene. (c) The sample is exposed to Si atoms at room temperature. (d) Si atoms are intercalated into the interface of graphene and Ir thin film after sample annealing.

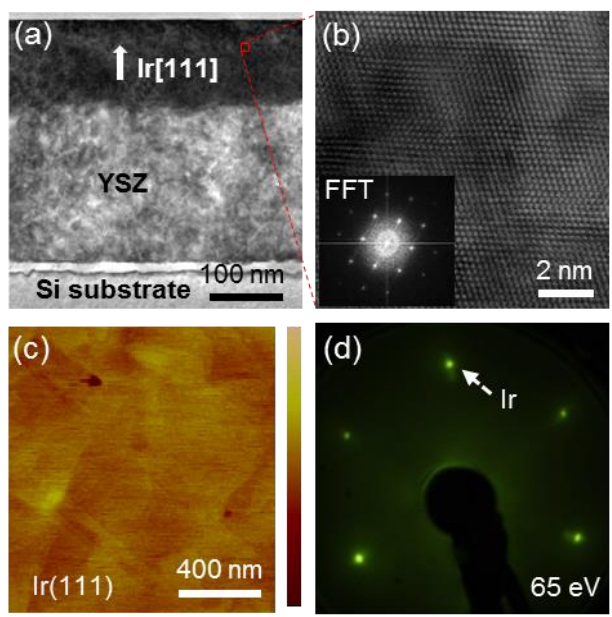

Figure 2. Structural and morphological characterization of Ir(111)/YSZ/Si(111) sample. (a) Cross-sectional TEM image of the heterostructures. (b) High-resolution TEM image taken at the cross-section of Ir film shows highly ordered structure of Ir films. Inset shows the corresponding FFT pattern. (c) AFM morphologies (top view) of the Ir(111) films before surface cleaning in ultrahigh vacuum. The color scale is 7.5 nm. (d) LEED pattern (obtained at beam energy of 65 eV) of clean sample, showing a single-crystalline Ir(111) facet.

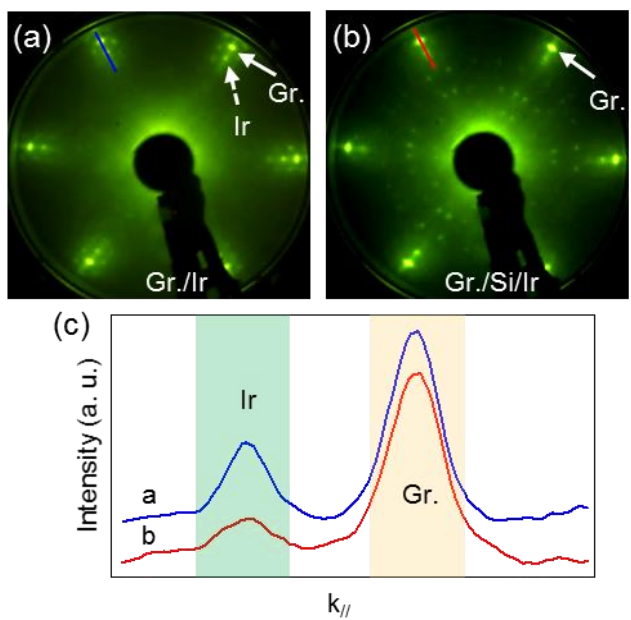

**Figure 3.** LEED pattern of the as-grown graphene on Ir(111) thin films before (a) and after (b) silicon intercalation. Spots corresponding to Ir(111) lattice and graphene adlayer are indicated by dash-line and solid-line arrows, respectively. (c) Intensity profiles of LEED patterns of graphene on Ir before and after Si intercalation across the solid lines in (a) and (b), respectively. For clarity, the intensities were normalized by the intensity for the graphene spots. LEED pictures in (a) and (b) were taken at beam energy of 65 eV.

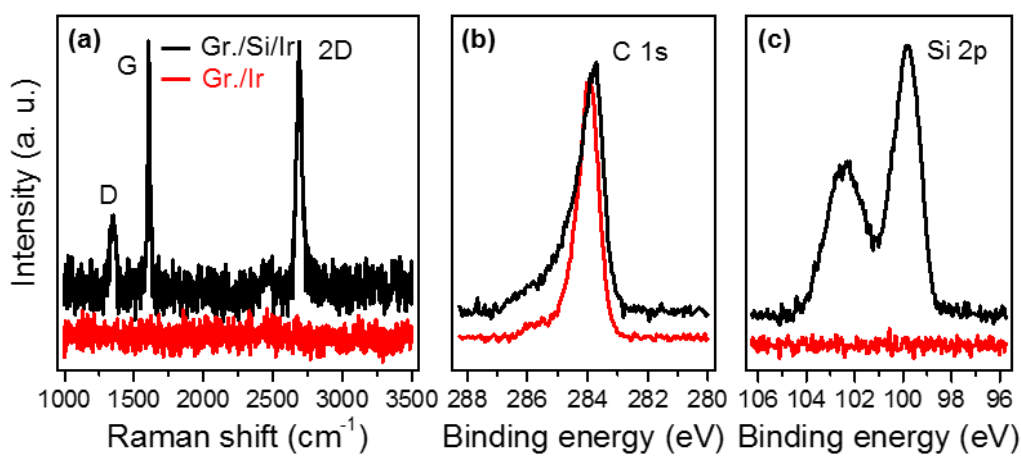

**Figure 4.** (a) Raman spectra of the graphene on Ir(111) thin films before (red, bottom curves) and after (black, top curves) Si intercalation. The dominant Raman features of graphene, G and 2D peaks, are visible in the Si-intercalated sample but invisible in the pre-intercalation sample. XPS of the C 1s (b) and Si 2p (c) before and after Si intercalation.

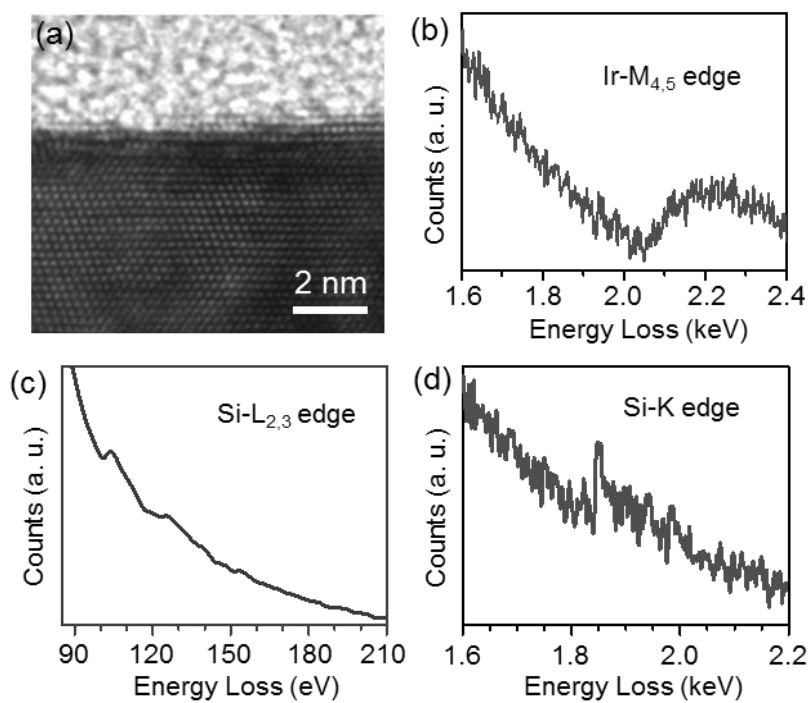

**Figure 5.** Interface of graphene/Si/Ir. (a) High-resolution cross-sectional TEM images taken at the interface of the silicon-intercalated graphene on Ir(111) thin films. (b) EEL spectrum of Ir-$M_{4,5}$ edge taken in the Ir film. EEL spectra of Si-$L_{2,3}$ edge (c) and Si-K edge (d), both taken at the interface.

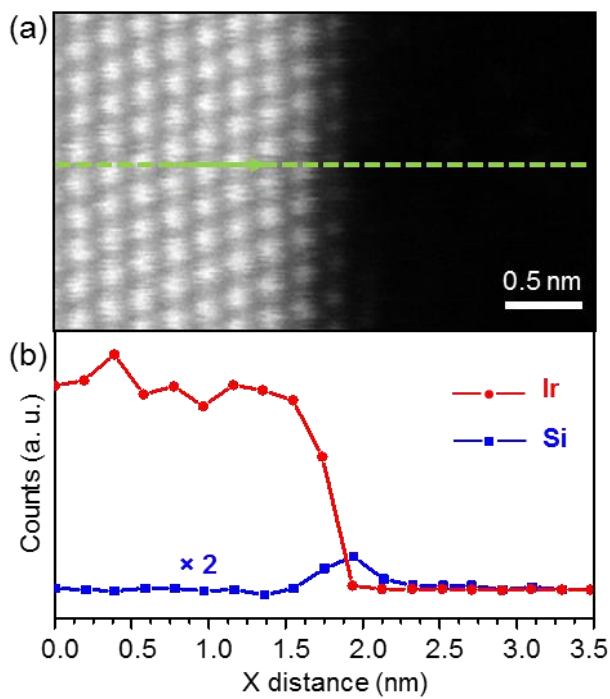

**Figure 6.** (a) High-resolution cross-sectional STEM images taken at the interface of graphene/Si/Ir thin film. (b) An EELS line scan across the interface as indicated by the light green dash line in (a). The arrow represents the line scan direction.

**Single-crystalline transition metal films** are ideal playing fields for the epitaxial growth of graphene and graphene-base materials. Graphene-silicon layered structures were successfully constructed on Ir(111) thin film on Si substrate with an yttria-stabilized zirconia buffer layer via intercalation approach. Such hetero-layered structures are compatible with current Si-based microelectronic technique, showing high promise for applications in future micro- and nano-electronic devices.

**Key word:** graphene-silicon, layered structures, Ir(111), single-crystalline thin film

Y. D. Que, Y. Zhang, Y. L. Wang, L. Huang, W. Y. Xu, J. Tao, L. J. Wu, Y. M. Zhu, K. Kim, M. Weinl, M. Schreck, C. M. Shen, S. X. Du, Y. Q. Liu, and H.-J. Gao[*]

**Graphene-Silicon Layered Structures on Single-crystalline Ir(111) Thin Film**

ToC figure

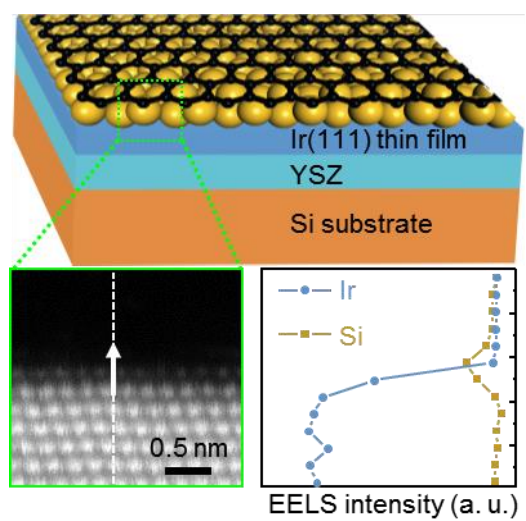